\documentclass[reprint,amsmath,amssymb,prb]{revtex4-2}

\usepackage{graphicx}
\usepackage{dcolumn}
\usepackage{bm}
\usepackage{xcolor}

\begin{document}

\title{Intrinsic Spatial Resolution Limit in\\ Analyzer-Based X-Ray Phase Contrast
Imaging Technique}

\author{Marcelo G. H\"{o}nnicke}
\email{marcelo.honnicke@unila.edu.br}
\affiliation{Universidade Federal da Integração Latino-Americana\\ 85867-970, Foz do Iguaçu, PR, Brazil}

\author{S\'{e}rgio L. Morelh\~{a}o}
\email{morelhao@if.usp.br}
\affiliation{Instituto de Física, Universidade de São Paulo\\ 05508-090, São Paulo, SP, Brazil}

\date{\today}
\begin{abstract}
Dynamical diffraction effects always play a role when working with perfect single crystals. The penetration of X-rays respect to the surface normal during diffraction (extinction depth, $1/\sigma_e$) in perfect single crystals does not have a constant value. The value changes for different angular positions on the crystal diffraction condition. For higher X-ray energies this value can change from few micrometers to tens of millimeters for each different crystal
angular position in the small angular range of the diffraction condition. This effect may spread a single point in
the object (sample) as a line in the image detector, especially if the crystal is set (or if the sample angularly
deviates the beam) at lower diffraction angle positions, where the surface component of X-ray penetration can
achieve huge values. Then, for imaging experiments where the dynamical diffraction occurs, such intrinsic
property can affect the image resolution. We have modeled and experimentally checked such a dynamical diffraction
property using, as example, an Analyzer-based X-ray phase contrast imaging setup (ABI) at two different
X-ray energies: 10.7\,keV and 18\,keV. The results show that our theoretical model is consistent with the measured
results. For higher energies the blur effect is enhanced and intrinsically limits the image spatial resolution.
\end{abstract}


\maketitle

\section{Introduction}

Dynamical diffraction effects always play a hole when working with perfect and nearly perfect single crystals (strained due to stress crystals). Within the dynamical condition, the penetration of X-rays respect to the surface normal during diffraction (extinction depth) in perfect single crystals does not have a constant value \cite{zp78,la98,aa01,mh05,mh08a,sm11}. The value changes for different angular positions on the crystal diffraction condition. For higher X-ray energies this value can change from few micrometers to tens of millimeters for each different crystal angular position in the small angular range of the diffraction condition \cite{mh08a}. Such an effect can be minimized for nearly perfect single crystals, since the strain due to stress, strongly affects the extinction \cite{aa01,bt76}.

Then, for imaging experiments, when dynamical diffraction occurs \cite{td95,sm10,wz14,oc17,wl01,gj08,as17,wh18}, the variable extinction depth may spread a single point in the object (sample) as a line in the image detector. This spoils the image resolution especially if the crystal (or portion of the sample) is set (or it is) at the lower diffraction angle position on its diffraction profile (rocking curve), where the surface component of X-ray extinction can achieve huge values. Note that, very often, in imaging experiments when dynamical diffraction occurs, the extinction depth is considered, theoretically, to have a constant value since the major part of the works take use of the extinction length (or Pendell{\"{o}}sung length, $\Lambda_B$) which presents a constant value \cite{wh18,vk01,yn04}.

\begin{figure*}
\includegraphics[width=6.75in]{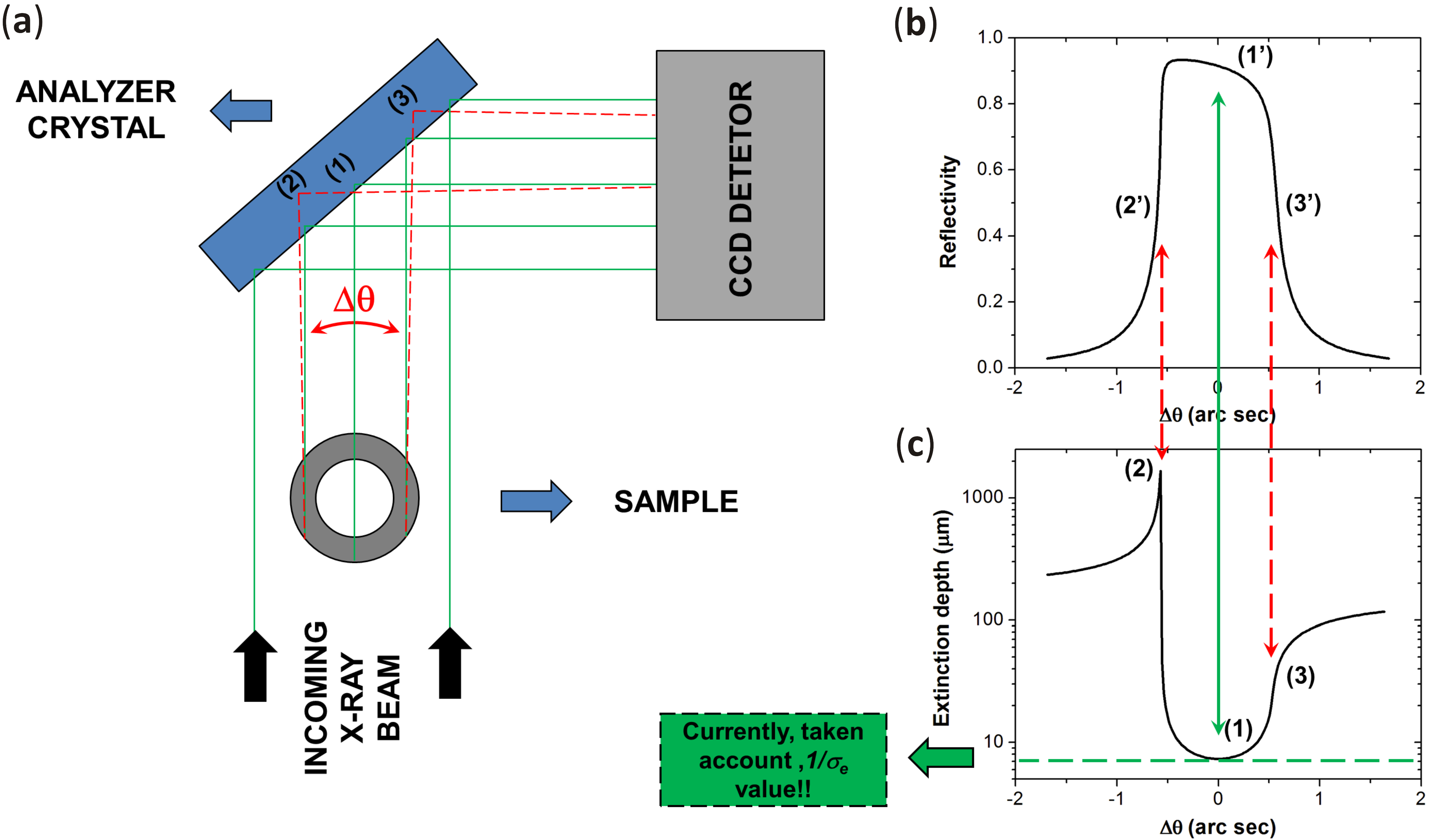}
\caption{\label{fig1}(a) Schematic representation of an analyzer-based X-ray phase contrast imaging (ABI) for a plane and monochromatic wave beam. The sample slightly deviates, angularly ($\Delta\theta$), the portion of the X-ray beam that cross the sample. Such deviation can be seen as an angular scan of the beam by the analyzer crystal. As an example, only three different positions (1–3) are represented in the figure. In our model to simulate the images, around 30 different positions were considered (30 positions on the beam cross section that intersects the sample). (b) The portion of the X-ray beam crossing the sample angularly scan the analyzer crystal as a rocking curve where, in this example, only three different angular positions are shown ($1^\prime$-$3^\prime$) The different angular positions correspond to different extinction depth ($1/\sigma_e$) on the analyzer crystal (1–3) (c). The variable extinction depth is responsible for changing (blurring) the projected image.}
\end{figure*}

In this work, the variable extinction depth effect is theoretically and experimentally explored, using as an example, an analyzer-based X-ray phase contrast imaging setup (ABI) \cite{mh07,mh08b}, Fig.~\ref{fig1}, with symmetrically-cut perfect single crystals at two different X-ray energies (10.7 keV and 18 keV). Theoretical studies were modelled by simulating the ABI images of a 300\,$\mu$m polyamide wire. Two different approaches were employed in the simulations: \emph{(i)} analyzer crystal for a plane and monochromatic X-ray wave beam; and \emph{(ii)} non-dispersive double crystal setup. For the modelling validation, the simulated images were compared with measured ones taken from a real 300\,$\mu$m polyamide wire. It is good to mention here, that quantitative analyzer based X-ray phase contrast imaging have been widely explored in the literature \cite{vk01,yn04,td96,mk07,ab07,am07,hz07,mh12,km14} and where, when mentioned, the extinction depth, based on the extinction length \cite{vk01,yn04} is considered to have a constant value.

\section{Dynamical diffraction and extinction depth}

For determining the X-ray penetration in single crystals in (extinction depth) and out (penetration depth) of the diffraction condition, we need to explore the Dynamical Theory of X-ray diffraction for plane and monochromatic X-ray wave beam approximation. Then, the penetration depth can be defined as \cite{zp78}:
\begin{equation}\label{eq1}
    \frac{1}{\sigma} = \frac{\gamma_0}{\mu}
\end{equation}
where $\sigma$ is the attenuation factor, $\mu$ is the linear attenuation coefficient, and $\gamma_0$ is the direction cosine of the incident angle respect to the crystal surface normal direction.

For an easy representation of the extinction depth $1/\sigma_e$, where $\sigma_e$ is the extinction factor, we firstly define the $y$ scale (angularly dependent):
\begin{equation}\label{eq2}
    y = \frac{-\chi_0+\left(1 - \frac{|\chi_0|}{2} \right)(\theta-\theta_0)\sin2\theta}{C |\chi_h|}
\end{equation}
where $\chi_0$ and $\chi_h$ are the polarizabilities, $\theta_0$ is the diffraction angle, $\theta$ is the Bragg angle and $C$ is the polarization factor. In the $y$ scale, the angular range where the diffraction occurs, is divided in three different regions: $y > 1$ (I) (maximum wavefield amplitude on the atomic planes), $1 > y > -1$ (II) (maximum reflectivity region) and $y < -1$ (III) (maximum wavefield amplitude between the atomic planes) \cite{sm16}. Then, the extinction depth for the adjacent I and III regions of the maximum reflectivity are given by:
\begin{equation}\label{eq3}
    \frac{1}{\sigma_e^{\rm I,III}}=\frac{\gamma_0}{\mu\left| \frac{y-\varepsilon}{\sqrt{y^2 - 1}}\right|}
\end{equation}
where $\varepsilon$ is the dielectric constant. And for the maximum reflectivity region,
\begin{equation}\label{eq4}
    \frac{1}{\sigma_e^{\rm II}}=\frac{\gamma_0}{\mu\left| \frac{C|\chi_{hr}|}{|\chi_{0i}|} \sqrt{1-y^2}\left(1+\frac{b^2}{8(1-y^2)}\right)\right|}
\end{equation}
where $\chi_{hr}$ and $\chi_{0i}$ are the real and imaginary parts of the polarizabilities $\chi_{h}$ and $\chi_{0}$, respectively. The extinction depth for $y = \pm1$ is 
\begin{equation}\label{eq5}
    \frac{1}{\sigma_e^{\pm1}}=\frac{\gamma_0}{\mu\left| \frac{C|\chi_{hr}|}{|\chi_{0i}|} (1\mp\varepsilon)\right|}\,.
\end{equation}
From equations (\ref{eq2}) to (\ref{eq5})
one can plot the extinction depth \emph{versus} angle as shown in Fig.~\ref{fig1}(c). Note that, the extinction depth limit for large $|y|$ is the penetration depth $1/\sigma$. For imaging experiments when dynamical diffraction occurs, the extinction depth is very often taken as the extinction length and considered, theoretically, to have a constant value \cite{wh18,vk01,yn04}, that is 
\begin{equation}\label{eq6}
    \frac{1}{\sigma_e^{\rm constant}}=\frac{\Lambda_B}{\gamma_0}\,.
\end{equation}
If one takes $y = 0$ in equation (\ref{eq4}), nearly the same value given by equation (\ref{eq6}) is obtained, as schematically indicated (green-dashed line) in Fig.~\ref{fig1}(c).

\section{Imaging simulation methods}

Computer codes for two different approaches were implemented in Matlab or Octave. Initially, a monochromatic X-ray plane wave was assumed as incident beam. Then, to compare with experimental results, the approach was improved by considering a non-dispersive double crystal setup as incident beam conditioner. Both approaches were applied for Bragg case ABI in either setups: 333 reflection in a Si(111) crystal at 10.7 keV; and 444 reflection in a Si(111) crystal at 18 keV. 

\subsection{Monochromatic X-ray plane wave}

\begin{figure*}
\includegraphics[width=6.75in]{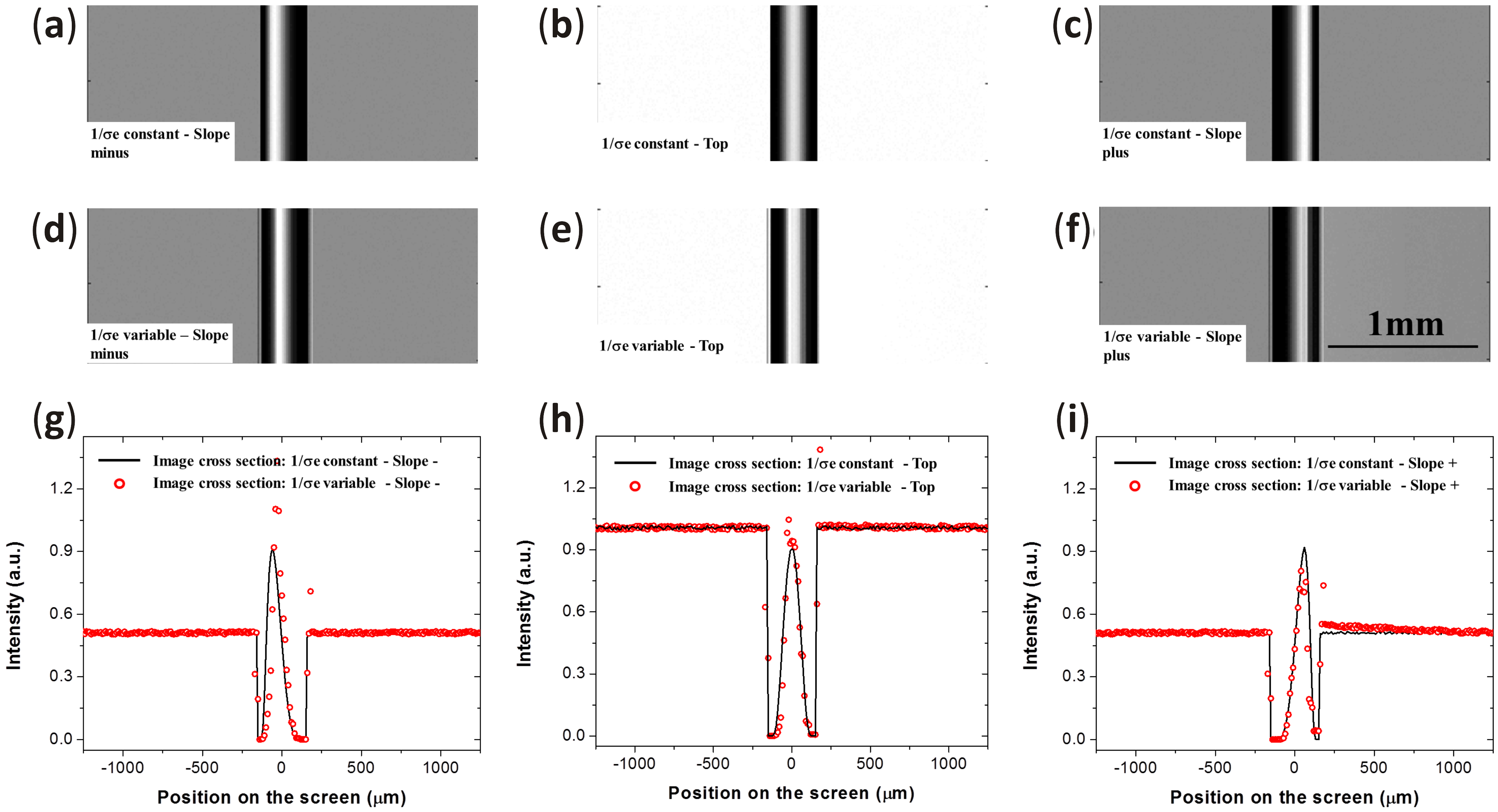}
\caption{\label{fig2} Simulated analyzer-based X-ray phase contrast images (ABI) of a 300\,$\mu$m thick polyamide wire, for a plane and monochromatic X-ray wave beam and Si 333
analyzer crystal at 10.7\,keV. (a–c) Slope minus, top and slope plus images for $1/\sigma_e$ constant. (d–f) Slope minus, top and slope plus images for $1/\sigma_e$ variable. (g–i)
Image cross sections. Solid black lines: $1/\sigma_e$ constant. Open red circles: $1/\sigma_e$ variable.}
\end{figure*}

\begin{figure*}
\includegraphics[width=6.75in]{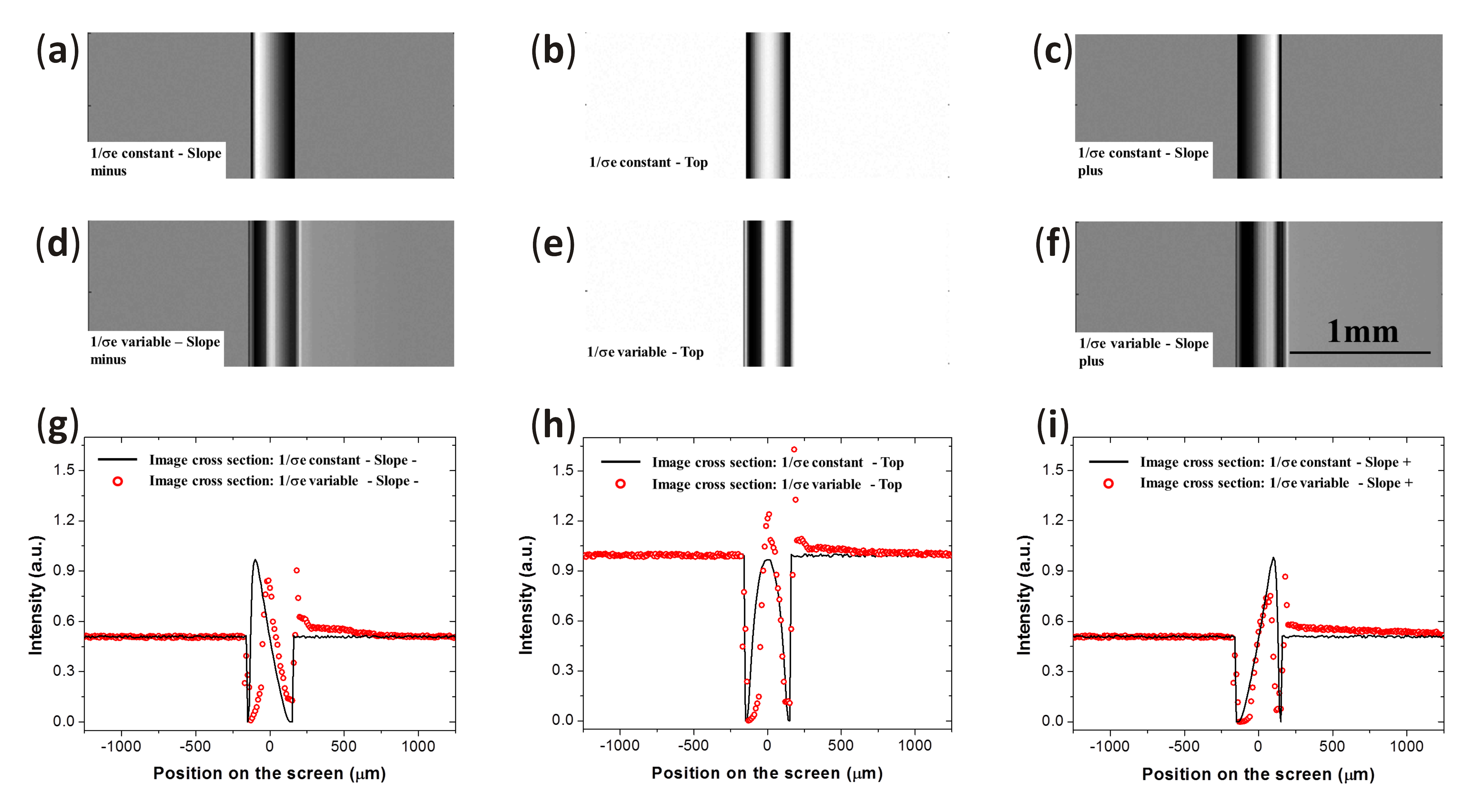}
\caption{\label{fig3} 
Simulated analyzer-based X-ray phase contrast images (ABI) of a 300\,$\mu$m thick polyamide wire, for a plane and monochromatic X-ray wave beam and Si 444
analyzer crystal at 18\,keV. (a–c) Slope minus, top and slope plus images for $1/\sigma_e$ constant. (d–f) Slope minus, top and slope plus images for $1/\sigma_e$ variable. (g–i)
Image cross sections. Solid black lines: $1/\sigma_e$ constant. Open red circles: $1/\sigma_e$ variable.}
\end{figure*}

For this approach we firstly calculate the extinction depth and the Darwin-Prins curves of Si 333 at 10.7 keV and Si 444 at 18 keV for symmetric Bragg case, as for instance the curves for Si 333 at 10.7 keV in Figs.~\ref{fig1}(b) and (c). These curves were stored in the database. After that, we have simulated the angular deviations of the X-ray beam, including the X-ray beam attenuation, for a $300\,\mu$m polyamide wire. This simulation was done with a resolution of $10\,\mu$m, i.e., the portion of the beam cross section intercepting the sample was striped into 30 sections. As each strip is characterized by one angular deviation ($\Delta\theta$) of the beam and, each angular deviation can be seen as an angular scan of the beam by the analyzer crystal, the angular deviation for each strip was stored in the database and then matched by the corresponding values of extinction depth and reflectivity. In other words, reflectivity and extinction depth values were attributed to each sample strip. The intensity of each strip registered in a two dimensional detector, and  spread in different areas corresponding to different ($1/\sigma_e$)\,$\gamma_0$, Fig.~\ref{fig1}(a). The intensity in each different detector area is weighted by the maximum reflectivity value of each sample strip. Also the spread beam has an exponential decay over its cross section for each sample strip. This information is stored in a single image matrix and summed up for each one of the image strip. The final image results joined with their cross sections are shown in Fig.~\ref{fig2} for Si 333 at 10.7 keV and in Fig.~\ref{fig3} for Si 444 at 18 keV for three different angular positions on the analyzer crystal: slope minus, top, and slope plus, corresponding respectively to positions 2, 1 and 3 in Fig. ~\ref{fig1}(a). The results are compared with simulated images taken for constant ($1/\sigma_e$) values. Strong differences were detected for both cases. However, these simulations are for an ideal case (plane and monochromatic X-ray beam). To be more realists we carried on a similar approach for a non-dispersive double crystal setup for symmetric Bragg case that is very often used for ABI applications.

\subsection{Non-dispersive double crystal setup}

\begin{figure*}
\includegraphics[width=6.75in]{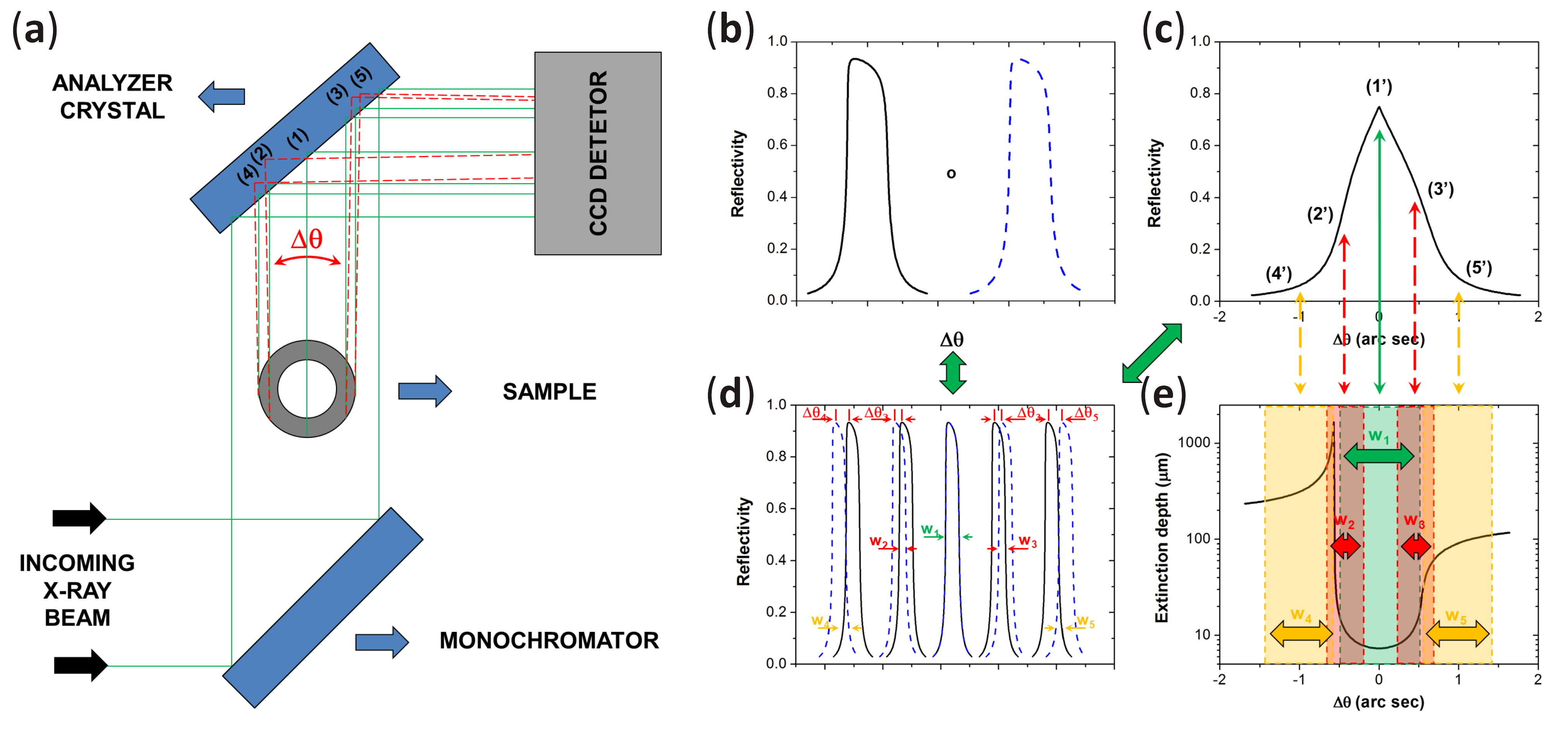}
\caption{\label{fig4}(a) Schematic representation of an analyzer-based X-ray phase contrast imaging (ABI) for a non-dispersive double crystal setup. As mentioned in Fig.~\ref{fig1}, the sample slightly deviates angularly ($\Delta\theta$), the portion of the X-ray beam which crosses the sample. Such deviation can be seen as an angular scan of the beam by the analyzer crystal. As an example, only five different positions (1–5) are represented in the figure. In our model to simulate the images, around 30 different positions were considered on the beam cross section intercepting the sample. (b) For non-dispersive double crystal setup, the portion of the beam deviated by the sample is angularly scanning the analyzer crystal as represented by a convolution of two Darwin-Prins curves, corresponding to the first (monochromator) and second (analyzer) crystals. (c) It results in a triangular-type of diffraction profile (rocking curve) where, as an example, five different angular positions ($1^\prime$-$5^\prime$) are indicated. (d) Schematic representation of the correlation process. For each different angular deviation (defined by different $\Delta\theta$s), a different angular width ($w_1$ to $w_5$) is restricted by the two crystals. The different widths are used to average (Gaussian normalized) the different extinction depths ($1/\sigma_e$), as indicated by  the different angular stripes.}
\end{figure*}

\begin{figure*}
\includegraphics[width=6.75in]{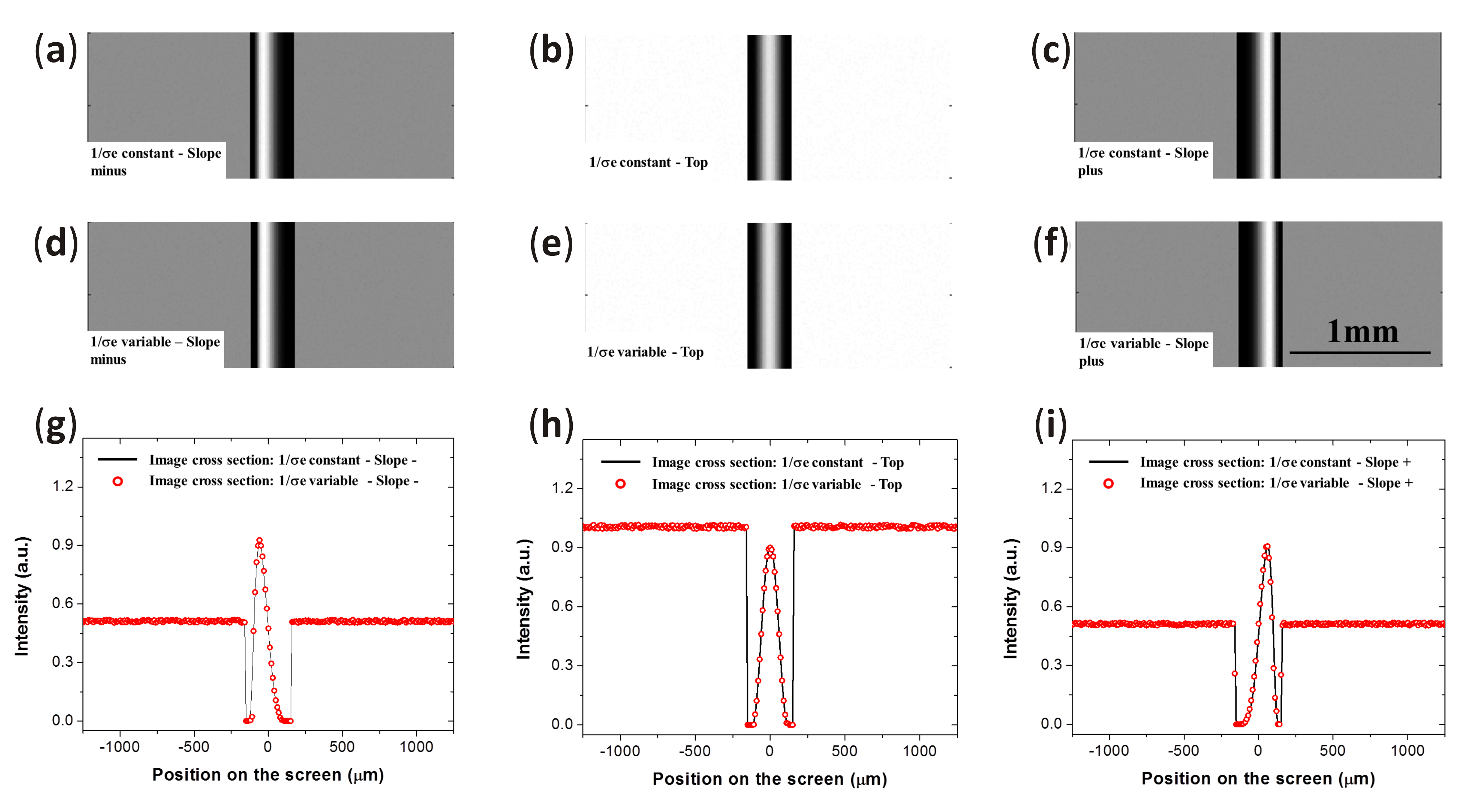}
\caption{\label{fig5} 
Simulated analyzer-based X-ray phase contrast images (ABI) of a 300\,$mu$m thick polyamide wire, for a non-dispersive double crystal setup and Si 333 analyzer
crystal at 10.7\,keV. (a–c) Slope minus, top and slope plus images for $1/\sigma_e$ constant. (d–f) Slope minus, top and slope plus images for $1/\sigma_e$ variable. (g–i) Image cross
sections. Solid black lines: $1/\sigma_e$ constant. Open red circles: $1/\sigma_e$ variable. For this approach at lower X-ray energies (10.7\,keV) any difference was detected among
constant and variable $1/\sigma_e$.}
\end{figure*}

\begin{figure*}
\includegraphics[width=6.75in]{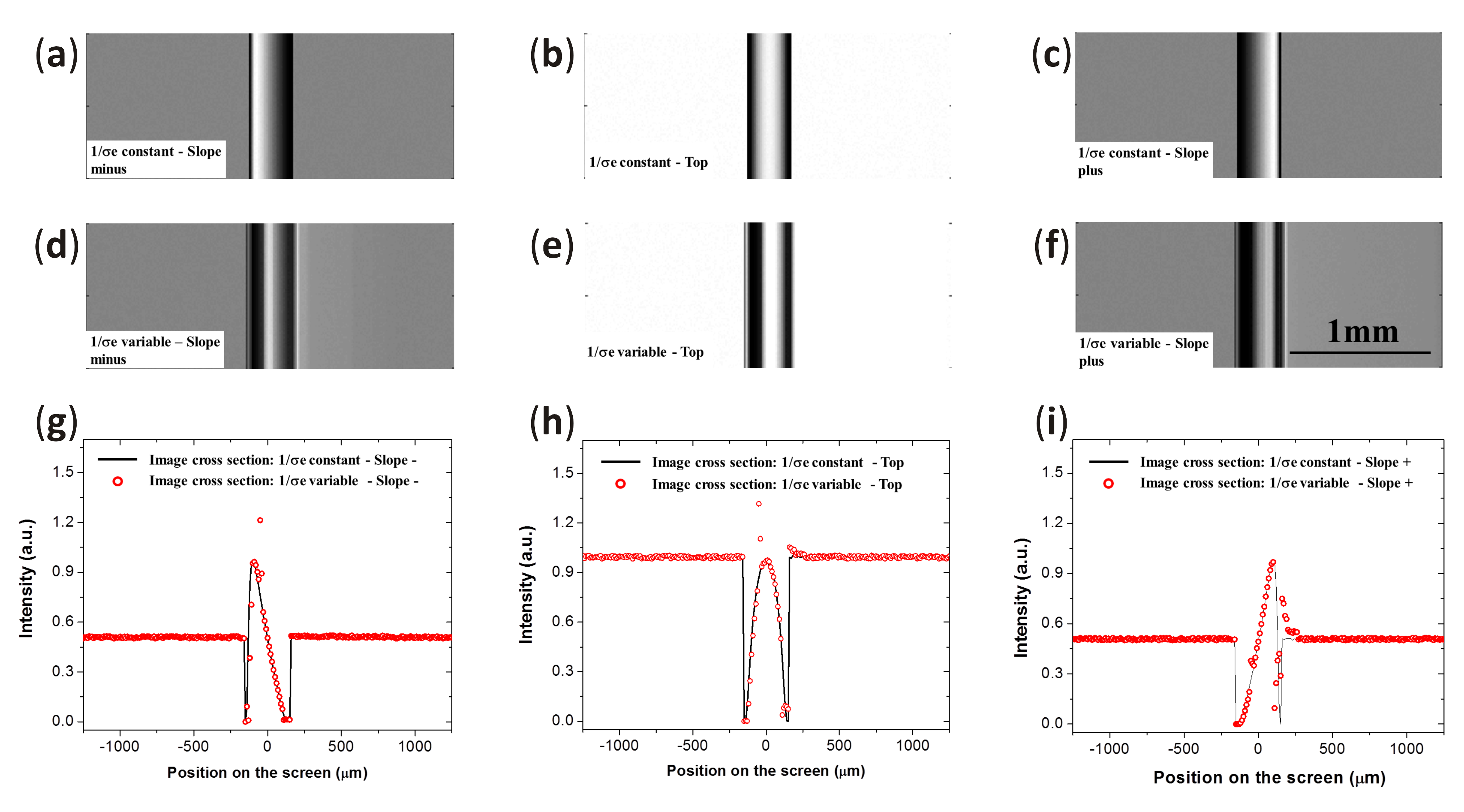}
\caption{\label{fig6}
Simulated analyzer-based X-ray phase contrast images (ABI) of a 300\,$\mu$m thick polyamide wire, for a non-dispersive double crystal setup and Si 444 analyzer
crystal at 18\,keV. (a–c) Slope minus, top and slope plus images for $1/\sigma_e$ constant. (d–f) Slope minus, top and slope plus images for $1/\sigma_e$ variable. (g–i) Image cross
sections. Solid black lines: $1/\sigma_e$ constant. Open red circles: $1/\sigma_e$ variable. For this approach at higher X-ray energies (18\,keV) differences were detected among
constant and variable $1/\sigma_e$.}
\end{figure*}

This setup is used for analyzer-based X-ray phase contrast imaging. The setup modeled here, Fig.~\ref{fig4}, is for a Bragg case and symmetrically cut
crystals. For this approach we firstly had to simulate the analyzer
rocking curve Fig.~\ref{fig4}(c), which now is a correlation between the Darwin-Prins curve of the first crystal (monochromator) and the Darwin-Prins curve of the second crystal (analyzer), Fig.~\ref{fig4}(b,d). The reflectivity curves of Si 333 at 10.7 keV and Si 444 at 18 keV for symmetric Bragg case were stored in the database. For the extinction depth values since each sample strip is characterized by one angular deviation ($\Delta\theta$) of the beam and, each angular deviation can be seen as an angular scan of the beam by the analyzer crystal, we can look for the schematic representation of the correlation procedure, Fig.~\ref{fig4}(d), and consider that, for each angular deviation there is a range of $1/\sigma_e$ limited by the width $w$, Fig.~\ref{fig4}(e). This gets the $1/\sigma_e$ profile smoother. Such an average procedure was modulated by a Gaussian profile. Then, the averaged $1/\sigma_e$ values for each angular deviation (or sample strip) was also stored in the database. Again, we have simulated the angular deviations of the X-ray beam, including the X-ray beam attenuation, for a $300\,\mu$m polyamide wire with the same parameters described in the previous sub-section. The stored angular deviations for each different sample strip are then matched with the corresponding averaged $1/\sigma_e$ and reflectivity values. Again, a reflectivity value and an extinction depth value are attributed to each sample strip. The intensity registered in a two dimensional detector, for each sample strip, is then spread in different areas corresponding to different averaged $(1/\sigma_e)\,\gamma_0$, Fig.~\ref{fig4}(a). The intensity in each area is normalized by the maximum correlated reflectivity value of each sample strip. As in the previous subsection, the spread beam has an exponential decay over its cross section for each sample strip. This information is stored in a single image matrix and summed up for each one of the image strip. The final image results joined with their cross sections are shown in Fig.~\ref{fig5} for Si 333 at 10.7\,keV and in Fig.~\ref{fig6} for Si 444 at 18\,keV for three different angular positions on the analyzer crystal: slope minus, top, and slope plus, corresponding respectivly to positions 2, 1 and 3 in Fig.~\ref{fig4}(a). The results are compared with simulated images taken for constant $1/\sigma_e$ values. For the lower energy, 10.7\,keV, no differences could be seen. Differences were detected only for the higher energy, 18\,keV. We can still try to estimate the blurring as function of energy although the blur depends, among other factors, of $(1/\sigma_e)\,\gamma_0$ and of the rocking curve width that changes from one diffraction plane to other, even for the same X-ray energy. The changes in both quantities strongly affect the blur sensitivity. Then, we simulated another set of ABIs for an X-ray beam energy of 14 keV with a non-dispersive double crystal Si 333 setup. Therefore, we have simulated ABIs at three different energies in order to estimate, in a graph, the contribution to the blurring as function of the X-ray beam energy. To do the graphics, first we need to quantify the blurring. For that we defined the Relative Image Blur 
\begin{equation}\label{eq7}
    RIB = \frac{\left(dI/dx\right)_{{1/\sigma_e} {\rm -variable}}}{\left(dI/dx\right)_{{1/\sigma_e} {\rm -constant}}}
\end{equation}

\begin{figure}
\includegraphics[width=3.25in]{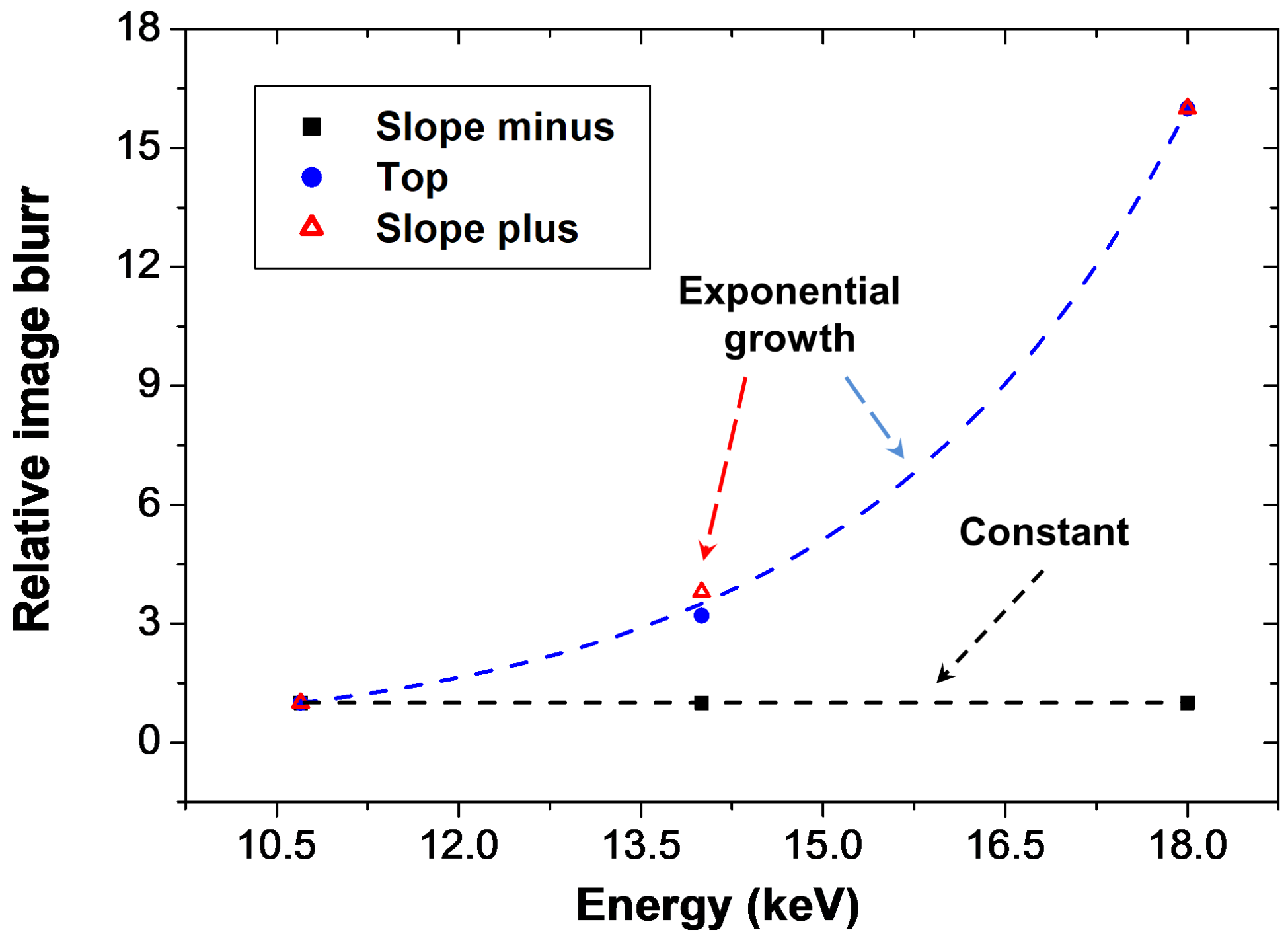}
\caption{\label{fig7} Relative image blur (RIB) versus X-ray beam energy calculated from the acquired simulated images for the non-dispersive double crystal setup for three different X-ray beam energies (10.7\,keV and Si 333, 14\,keV and Si 333 and 18\,keV and Si444) and three different angular positions on the analyzer crystal (slope minus, top and slope plus, which correspond to positions 2, 1 and 3, respectively in Fig.~\ref{fig4}(a). The RIB was determined across the left sample edge.}
\end{figure}

\noindent where $(dI/dx)_{{1/\sigma_e} {\rm -variable}}$ is the derivative of the intensity with respect to the position on the area detector across the sample edge for variable $1/\sigma_e$
ABIs, while $(dI/dx)_{{1/\sigma_e} {\rm -constant}}$ is the derivative of the intensity with respect to
the position on the area detector across the sample edge for constant $1/\sigma_e$ ABIs. The RIB results are shown in Fig.~\ref{fig7} for ABIs simulated at the
three different angular positions of the analyzer crystal: slope minus,
top, and slope plus, corresponding to positions 2, 1 and 3 in Fig.~\ref{fig4}(a). For the top and slope minus angular positions there is a tendency to an exponential growth of RIB. However, as previously mentioned, this is an estimative. More exhaustive simulations on different diffraction planes and diffraction plane orders at different X-ray beam energies need to be carried out in order to predict such RIB behaviour. To check the validity of this theoretical approach (blur due
to the dynamical diffraction condition), simulated images are compared with experimental results as described in the next section.

\section{experimental results}

Experimental
images of a $300\,\mu$m polyamide wire were acquired with two different ABI
non-dispersive setups: Si 333 double crystal setup at 10.7 keV and Si
444 double crystal setup at 18 keV, described elsewhere \cite{mh07,mh08b}. However, since the detectors are different in both experimental cases, we had to include in our scripts blur due to the image detector instrumentation.

\begin{figure}
\includegraphics[width=3.25in]{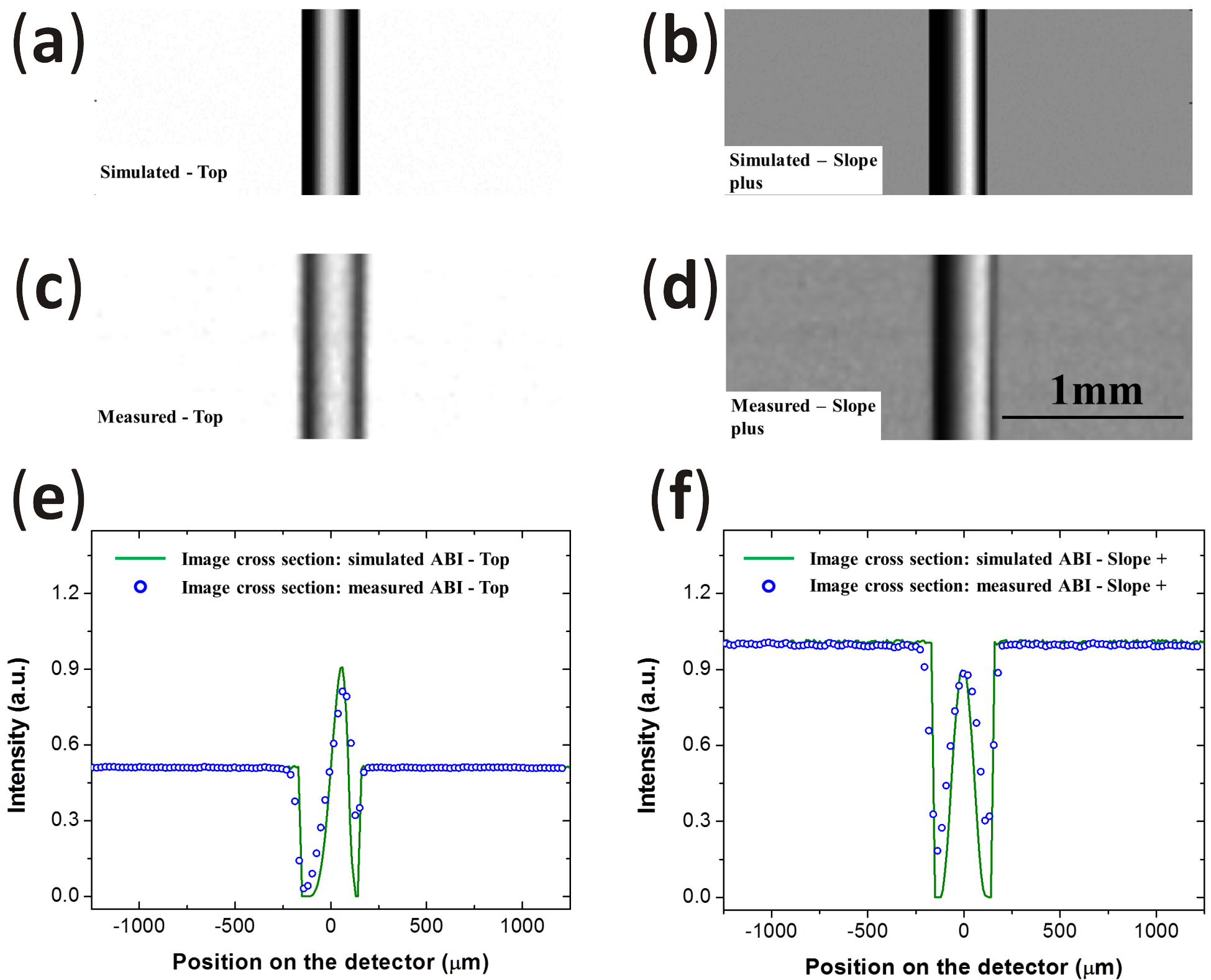}
\caption{\label{fig8} Experimental validation for a non-dispersive double crystal setup and Si 333 analyzer crystal at 10.7\,keV. (a–b) Top and slope plus simulated analyzer-based Xray phase contrast images (ABI) ($1/\sigma_e$ variable) of a 300\,$\mu$m thick polyamide wire. (c–d) Top and slope plus measured images. (e–f) Image cross sections. Solid green
lines: simulated ABIs. Open blue circles: measured ABIs.}
\end{figure}

For the 10.7\,keV ABI setup a direct conversion CCD detector with pixel size of $22.5 \times 22.5\,\mu{\rm m}^2$ was employed. As the theoretical pixel
size of our theoretical model is $10 \times 10\,\mu{\rm m}^2$, we included a $2\times2$ binning in our final simulated image. The simulated image results
including the detector contribution and the measured images joined with their image cross sections are shown in Fig.~\ref{fig8} for two different angular positions of the analyzer crystal: top ($1^\prime$) and slope plus ($3^\prime$) in Fig.~\ref{fig4}(c).

\begin{figure}
\includegraphics[width=3.25in]{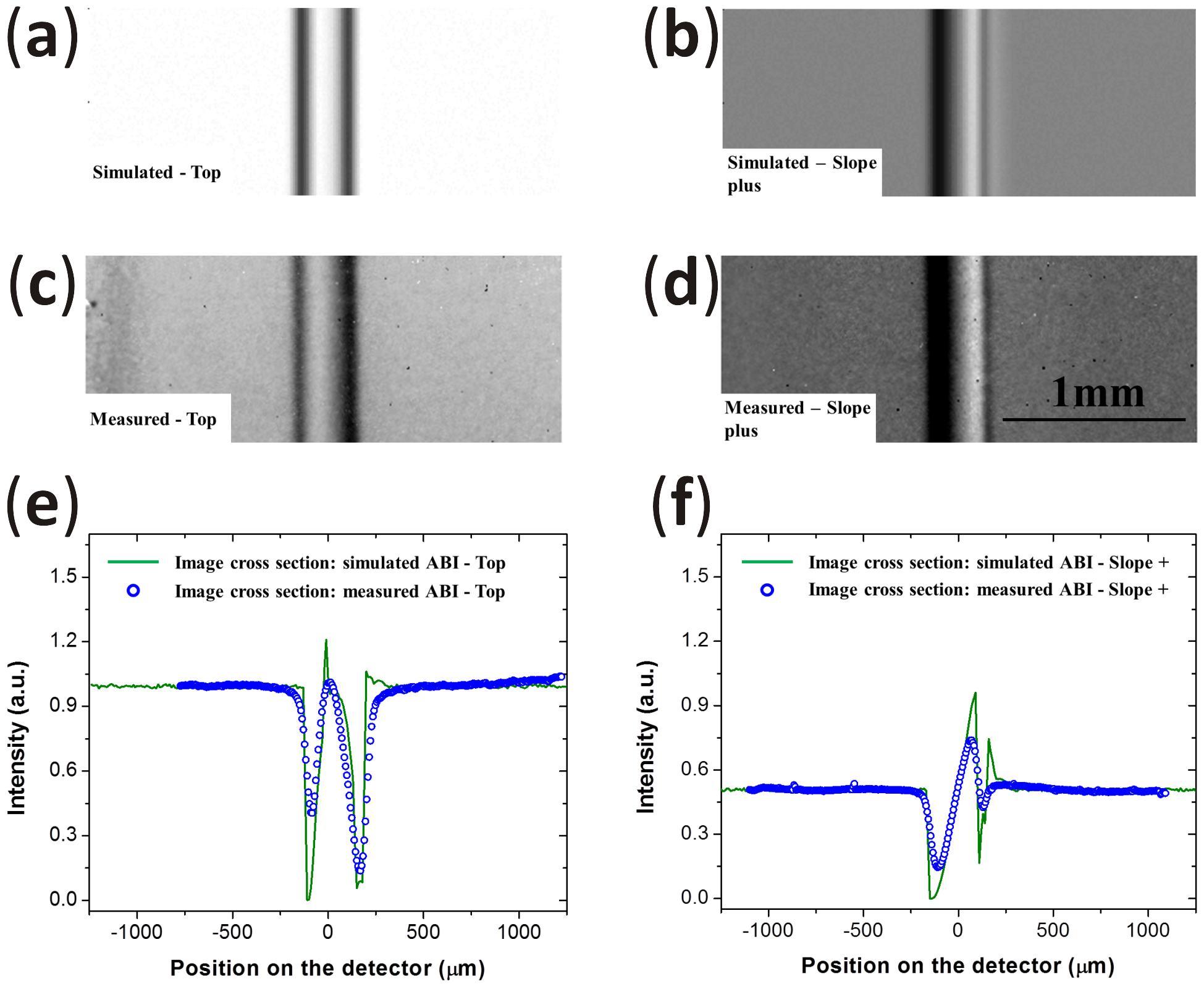}
\caption{\label{fig9} Experimental validation for a non-dispersive double crystal setup and Si 444 analyzer crystal at 18\,keV. (a–b) Top and slope plus simulated analyzer-based Xray phase contrast images (ABI) ($1/\sigma_e$ variable) of a 300\,$\mu$m thick polyamide wire. (c–d) Top and slope plus measured images. (e–f) Image cross sections. Solid green
lines: simulated ABIs. Open blue circles: measured ABIs.}
\end{figure}

For the 18 keV ABI setup, an indirect conversion CCD detector with pixel size of $3.5 \times 3.5\,\mu{\rm m}^2$ and a $2 \times 2$ binning was employed. The
calculated spatial resolution of this detector with this binning is $24 \times 24\,\mu{\rm m}^2$. This was included in our simulated images with a point
spread function (PSF) of $24\,\mu{\rm m}$. The simulated image results including the detector contribution and the measured images joined with their
image cross sections are shown in Fig.~\ref{fig9} for two different angular positions of the analyzer crystal: top ($1^\prime$) and slope plus ($3^\prime$) in Fig.~\ref{fig4}(c).

The results for both energies are consistent with the used theoretical approach. There are
slight differences between the simulated and measured images for the setup at lower energy (10.7 keV ABI non-dispersive double crystal setup), which is attributed to the narrower theoretical diffraction profile compared with the measured one. This can be easily adjusted. Other slight differences
were also found for the higher energy setup (18 keV ABI non-dispersive double crystal setup). In Fig.~\ref{fig9}(e), the measured image was not acquired exactly in the top of the analyzer crystal rocking curve, which can also be easily adjusted in the simulation. Another point are the peaks found in the middle Fig.~\ref{fig9}(e) and in the right side of Fig.~\ref{fig9}(f). These peaks appeared in the
simulated images and are correlated to the singularity in the extinction depth for the maximum value. However, in Fig.~\ref{fig9}(f), 
there is only a very slight tendency of the measured image to follow such a peak. We need to test our model with other image models in order to check if such peaks can be found. Since the singularity point is strongly sensible to stress due strain in the crystal, this value can easily be reduced by a factor of 5. Also for this specific point the divergence of the beam can play an important role, since the width of the $1/\sigma_e$ is angularly narrow at this position. The implementation of the divergence in
our code is envisaged.

\begin{figure}
\includegraphics[width=3.25in]{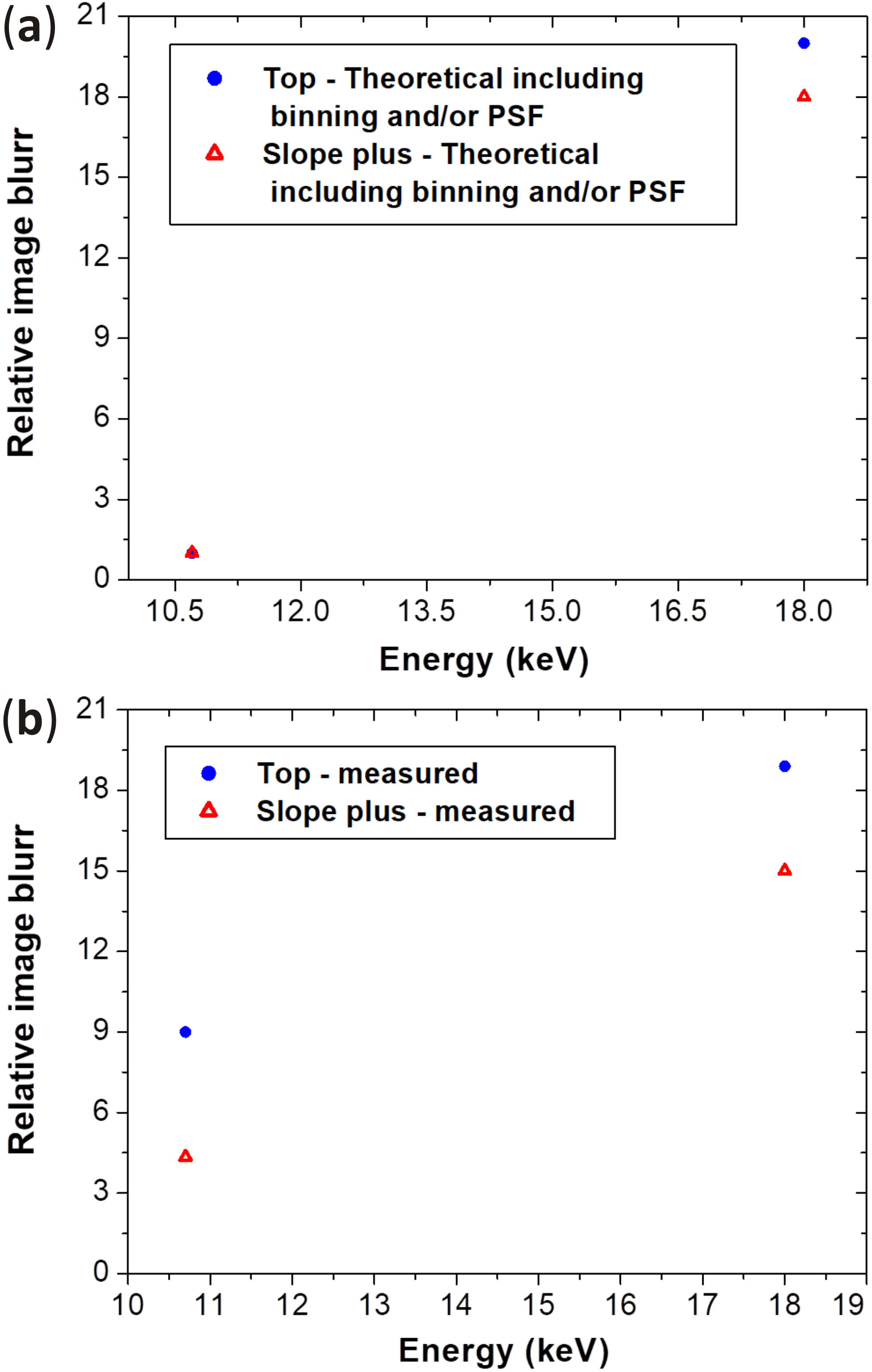}
\caption{\label{fig10} Relative image blur (RIB) versus X-ray beam energy for experiment validation purposes. (a) Calculated RIB for the non-dispersive double crystal setup for two different energy values (10.7\,keV and Si 333 and 18\,keV and Si444) including the detector binning and/or the detector point spread function (PSF) for two different angular positions on the analyzer crystal (top and slope plus, which correspond to 1 and 3, respectively in Fig.~\ref{fig4}(a). (b) Experimental RIB for the same parameters described in (a). The RIBs were determined across the left sample edge.}
\end{figure}

To be more quantitative, we calculated the RIB for the simulated ABIs including the detector contribution as well as for the measured ABIs. The results are shown in Fig.~\ref{fig10}. Significant differences were
found in the lower energies ABIs which, as mentioned in the previous paragraph, can be can be attributed to the narrower theoretical diffraction profile compared with the measured one.

\section{results and discussions}

We have modeled a dynamical diffraction based X-ray imaging experiment taking into account the variable extinction depth ($1/\sigma_e$). By the simulations we have shown, for an analyzer-based X-ray phase contrast imaging setup (ABI) and a plane and monochromatic X-ray
wave beam, that such dynamical diffraction property can severely blur the acquired images, especially at higher energies (18 keV). A more realistic simulation, based on a non-dispersive double crystal setup, including image detector contributions showed close agreement between the simulated and measured images for two different ABI nondispersive setups (Si 333 double crystal setup at 10.7 keV and Si 444 double crystal setup at 18 keV). Slight differences between the simulated and measured image cross sections were be attributed to: \emph{i)} narrower theoretical diffraction profile compared with the measured one; \emph{ii)} slight difference between the theoretical and measured rocking curve angular where the image was acquire; and \emph{iii)} the singularity in the
extinction depth for the maximum value ($1/\sigma_e$) which is responsible for small peaks in the higher energy (18 keV) ABIs. Since the singularity
in extinction depth is strongly sensible to stress due strain in the crystal, this effect can easily suppressed in experimental results. Also, for this
specific angular position the divergence of the beam can play an important role, since the width of the $1/\sigma_e$ is angularly narrow at this
position. The implementation of the divergence in our code, to better estimate this, is envisaged.

\bibliography{ISRLmanuscript}

\end{document}